\documentclass[epj]{svjour}
\usepackage{graphicx}

\usepackage{bm}
\usepackage{amssymb,amsmath}

\begin{document}
\title{Weak-coupling Treatment of Electronic (Anti-)Ferroelectricity in the
Extended Falicov-Kimball Model}
\author{Claudia Schneider, Gerd Czycholl}
\institute{Institute for Theoretical Physics, University of Bremen, D-28334
Bremen, Germany}
\date{\today}
\abstract{
We study the (spinless) Falicov-Kimball model extended by a finite band width
(hopping $t_f$) of the localized (f-) electrons in infinite dimensions 
in the weak-coupling limit of a
small local  interband Coulomb correlation $U$ for half filling. In the case
of overlapping conduction- and f-bands different kinds of ordered solutions
are possible, namely charge-density wave (CDW) order, 
electronic ferroelectricity
(EFE) and electronic antiferroelectricity (EAFE). 
The order parameters are calculated as
a function of the model parameters and of the temperature. There is a
first-order phase transition from the CDW-phase to the EFE- or EAFE-phase. The
total energy is calculated to determine the thermodynamically stable
solution. The quantum phase diagrams are calculated.  
}

 % end of PACS codes
 %end of abstract
%
\authorrunning{Schneider, Czycholl}

\titlerunning{(Anti-)Ferroelectricity in the Extended Falicov-Kimball Model}

\maketitle
\section{Introduction}
\label{sec:intro}
One of the simplest lattice models for strongly correlated electron
systems, the  Falicov-Kimball model (FKM) \cite{FalicovKimball69}, consists
of two types of spinless electrons, namely delocalized band
($c$-) electrons and localized $f$-electrons, and a local Coulomb (Hubbard)
interaction between $c$- and $f$-electron at the same site. The FKM was
originally introduced as a model
for metal-insulator and valence transitions\cite{FalicovKimball69}. It can
also be interpreted as a model for crystallization (identifying the
''heavy'' $f$-particles with the  nuclei, the $c$-particles with the
electrons and for $U<0$)\cite{KennedyLieb86}. Furthermore, the FKM
is of interest for
academic reasons, because it is the simplest non-trivial lattice model for
correlated electron systems, for which certain exact results are available;
for a recent review see Ref. \cite{FreericksZlatic03}.

A few years ago it has been suggested by Portengen {\it et
al.}\cite{PortengenSham96} that a novel ferroelectric state could be present
in the mixed-valence regime of the FKM. Whereas a
ferroelectric transition is usually connected with a structural phase
transition\cite{Kittel}, a purely electronic mechanism would lead to this
kind of ferroelectricity suggested to occur for the
FKM; it has, therefore, been termed ''electronic ferroelectricity''
(EFE)\cite{PortengenSham96}. The origin of EFE is a non-vanishing
excitonic expectation value $P_{cf}=\langle c^{\dagger}f\rangle$. In the
case of a vanishing hybridization between $c$-and $f$-electron states and a
vanishing electrical (optical) field driving inter-band transitions, the
existence of a finite $P_{cf} \neq 0$ is a kind of symmetry breaking, and if
the $f$- and $c$-states have different parity, the $P_{cf}$  causes a finite
electrical polarization without a driving electrical field, because of which
it is ''ferroelectricity''.     

On the other hand, from exact results available for the
FKM\cite{KennedyLieb86,BrandtSchmidt,BrandtMielsch} one knows that a charge
density wave (CDW) phase (''chess board phase'') exists at least for half
filling and in the symmetric case, and no evidence for EFE in the FKM 
is obtained by these exact treatments. But the possibility of
CDW ordering was not considered by Portengen {\it et
al.}\cite{PortengenSham96}. Later 
calculations\cite{Czycholl99,Farkasovsky99,Farkasovsky02,Sarasua02} could not confirm
the existence of EFE for the FKM. A divergence in the hybridization
susceptibility obtained in Ref.\cite{Zlatic2001} does not necessarily
mean that the ground state has spontaneous hybridization $P_{cf} \neq
0$\cite{Farkasovsky02}. 

More recently it has been shown by Batista {\it et
al.}\cite{Batista02,Batistaetal04} that the FKM extended by a direct
$f-f$-hopping $t_f$, in fact, has the EFE phases with a spontaneous
hybridization. Depending on the sign of $t_f$ a ferro- or an
antiferroelectric phase may exist. A CDW phase is also possible depending on
the relative position of $c$- and $f$-band and on the value of the Coulomb
(Hubbard) correlation $U$, and the quantum phase diagrams were obtained in
Refs. \cite{Batista02,Batistaetal04}. In this paper we also study the
spinless extended Falicov Kimball model (EFKM) 
suggested by Batista\cite{Batista02}.
Batista {\it el al.} studied this model in one and two dimensions in the
strong\cite{Batista02} and the intermediate coupling\cite{Batistaetal04}
limit. Here we study an infinite dimensional system in the weak coupling
limit. We also obtain electronic ferroelectricity (EFE), electronic
antiferroelectricity (EAFE) and CDW ordering. We calculate the dependence of
the order parameters on temperature and on the model parameters and
calculate the total energy. The resulting quantum phase diagram is
qualitatively very similar to that obtained previously in the intermediate
and strong coupling limit\cite{Batista02,Batistaetal04}. We also point out
that this EFKM and the  E(A)FE problem and phase is closely related to the
excitonic insulator phase discussed already about 40 years
ago\cite{Kohn67,Keldysh64,Zittartz67}.   

The paper is organized as follows. In Sect. \ref{sec:model} we describe the
EFKM and point out its connections with other standard models of correlated
electron systems and solid state theory. Section \ref{sec:approximation}
describes our weak-coupling approximation. The results are presented in
Section \ref{sec:results}; the $c$-and $f$-electron spectral functions are
calculated for different model parameters and order types, the (CDW, EFE,
EAFE) order parameters and the total energy for these phases 
are calculated as a function of the
model parameters and the temperature, and the complete quantum phase diagram is presented,
before the paper closes in Sect. \ref{sec:conclusion} with a short summary
and conclusion.

\section{Model}
\label{sec:model}

The extended Falicov-Kimball model (EFKM)\cite{Batista02} consists
of two types of spinless electrons, here denoted as 
$c$-and $f$-electrons, and a local Coulomb (Hubbard)
interaction $U$ between $c$- and $f$-electron at the same site. The EFKM
Hamiltonian reads:
\begin{eqnarray}
\label{eq:efkm-hR}
H = \sum_{\bf R}  \left(E_c c_{\bf R}^{\dagger}c_{\bf R} + 
E_f f_{\bf R}^{\dagger}f_{\bf R} + 
U c_{\bf R}^{\dagger}c_{\bf R}f_{\bf R}^{\dagger}f_{\bf R}\right.&&\nonumber\\
-  \left. \sum_{{\bf \Delta} n.n.} 
\left[t_c c_{\bf R + \Delta}^{\dagger}c_{\bf R} +
t_f f_{\bf R + \Delta}^{\dagger}f_{\bf R}\right]\right)&&\\
\label{eq:efkm-hk}
= \sum_{\bf k} \left(\varepsilon_c({\bf k}) c_{\bf k}^{\dagger}c_{\bf k} +
\varepsilon_f({\bf k}) f_{\bf k}^{\dagger}f_{\bf k}\right)
+ U \sum_{\bf R} c_{\bf R}^{\dagger}c_{\bf R}f_{\bf R}^{\dagger}f_{\bf R}&&
\end{eqnarray} 
Here $\bf R$ denotes the sites of a Bravais lattice, $\bf \Delta$ the nearest
neighbor lattice vectors, $\bf k$ the wave vectors from the first Brillouin
zone, $E_{c/f}$ are the on-site one-particle matrix elements (and thus the
band centers) of the $c/f$-electrons, 
and the usual nearest neighbor tight-binding assumption (of only
nearest neighbor intersite matrix elements $t_{c/f}$) 
has been made. Therefore, the
$c$- and $f$-electron dispersions are given by:
\begin{eqnarray}
\label{eq:dispersion}
\varepsilon_c({\bf k}) = E_c - \sum_{{\bf \Delta} n.n.} t_c e^{i \bf k
\Delta} \; \; , 
\varepsilon_f({\bf k}) = E_f - \sum_{{\bf \Delta} n.n.} t_f e^{i \bf k
\Delta}
\end{eqnarray}

Several standard models of solid state theory can be identified to be certain
limiting cases of this EFKM (\ref{eq:efkm-hR},\ref{eq:efkm-hk}). In the case
of a vanishing $f$-electron dispersion, i.e. $t_f = 0$, we recover, of
course, the standard spinless FKM\cite{FalicovKimball69,FreericksZlatic03}.
In the case of equal, degenerate $c$- and $f$-bands, 
i.e. $E_f = E_c$ and $t_f=t_c$, one can identify the $c$-electrons with the
spin-up and the $f$-electrons with the spin-down electrons and  obtains
the standard Hubbard model\cite{Hubbard}. 
A particle-hole transformation for one kind of
electrons, say the $f$-electrons, leads to an attractive interaction $-U$; 
then in the case $t_f = -t_c, E_c+U = -E_f$ the $c$-electron band and the
$f$-hole band are again degenerate, and identifying again 
the $c$-electrons with the spin-up electrons and the $f$-holes with the
spin-down electrons one has spin-degenerate fermions with an attractive
(short ranged, i.e. $\bf k$-independent) s-like interaction, i.e. the
BCS-model\cite{BCS}. 
Finally, if the $f$-electron band is interpreted as  valence
band and the $c$-electron band as conduction band (of one spin direction), 
one recovers the standard two-band model studied frequently in semiconductor
theory\cite{HaugKoch}, in particular to describe optical excitations 
(excitons etc.) under
the influence of the Coulomb interaction, only that here this Coulomb
interaction is local (short ranged). In this situation  the $f$-(valence)
band is narrower than the $c$-(conduction) band, i.e. $|t_f| < |t_c|$, and
the $f$- and 
$c$-states have usually a different parity, which in the simplest way can be
modelled by a different sign of $t_c$ and $t_f$ (i.e. $t_c > 0 , t_f < 0$).
In our opinion it is just this fact that so many interesting and important
models of solid state theory can be identified as limiting cases of the EFKM
(\ref{eq:efkm-hR},\ref{eq:efkm-hk}), which makes the EFKM very useful and
interesting at least for academic reasons (model studies).

\section{Approximation}
\label{sec:approximation}

We will use the following model assumptions: we measure
energies relative to the
origin of the $c$-band, i.e. we choose $E_c = 0$. Furthermore, we use a
semielliptical model density of states for the unperturbed $c$-band, 
i.e. we assume
\begin{eqnarray}
\label{eq:semodeldos}
\rho_{c0}(E) = \frac{1}{N} \sum_{\bf k} \delta(E - \varepsilon_c({\bf k}))
&=&
\frac{2}{\pi} \sqrt{1 - E^2} \\
&& \mbox{ for } -1 < E < 1 \nonumber
\end{eqnarray}
thereby choosing half the unperturbed $c$-band width as our energy unit.
This implies 
\begin{eqnarray}
\label{eq:fmodeldos}
\rho_{f0}(E) = \frac{1}{|t_f|}\rho_{c0}(\frac{E-E_f}{|t_f|})
\end{eqnarray}
Then we are left with three parameters: the position $E_f$ of the (center of
the) $f$-band relative to the (center of the) $c$-band, the (dimensionless)
$f$-electron hopping $t_f$ (i.e. the relation of the $f$- to the $c$-band
width), and the Coulomb (Hubbard or Falicov-Kimball) 
correlation $U$ between one $f$- and one $c$-electron at the same lattice
site.

To use the semielliptical model DOS (\ref{eq:semodeldos}) is a standard
model assumption introduced already more than 40 years ago by
Hubbard\cite{Hubbard}. It becomes exact for a Bethe lattice in the limit of
infinite coordination number. Compared to the Gaussian model DOS, which
becomes exact for a $d$-dimensional (hyper)cubic lattice in the limit of
infinite dimensions ($d \rightarrow \infty$)\cite{MetznerVollh}
 or coordination number, the semielliptical model DOS has the advantage that
the band width is finite and true band gaps can develop and 
that it has the squareroot band edge van Hove
singularities characteristic for three dimensional systems.

With these additional model assumptions we now apply the generalized
(unrestricted) Hartree-Fock approximation (HFA), which becomes correct in
the weak-coupling limit of small $U$. Within HFA the many-body (interaction)
part of the Hamiltonian (\ref{eq:efkm-hR},\ref{eq:efkm-hk}) is decoupled
according to
\begin{eqnarray}
\label{eq:HFA-decoupling}
c_{\bf R}^{\dagger}c_{\bf R}f_{\bf R}^{\dagger}f_{\bf R} &=& 
\langle c_{\bf R}^{\dagger} c_{\bf R}\rangle f_{\bf R}^{\dagger}f_{\bf R} + 
\langle f_{\bf R}^{\dagger}f_{\bf R}\rangle c_{\bf R}^{\dagger}c_{\bf R} \\
&-& 
\langle c_{\bf R}^{\dagger}f_{\bf R} \rangle f_{\bf R}^{\dagger}c_{\bf R} 
- \langle f_{\bf R}^{\dagger}c_{\bf R}\rangle c_{\bf R}^{\dagger}f_{\bf R}
\nonumber 
\end{eqnarray}  
It is an unrestricted HFA because we also allow for a decoupling with
respect to off-diagonal (excitonic) expectation values 
$\langle c_{\bf R}^{\dagger}f_{\bf R} \rangle$ and because we allow for a
positon-($\bf R$-)dependence of the expectation values 
$\langle c_{\bf R}^{\dagger} c_{\bf R}\rangle$, 
$\langle f_{\bf R}^{\dagger}f_{\bf R}\rangle$ and
$\langle c_{\bf R}^{\dagger}f_{\bf R} \rangle$. Within HFA the full
Hamiltonian (\ref{eq:efkm-hR}) is replaced by the effective one-particle
Hamiltonian
\begin{eqnarray} 
\label{eq:efkm-HFA}
H_{\mbox{\small eff}} &=& \sum_{\bf R}  
\left(\tilde{E}_{c {\bf R}} c_{\bf R}^{\dagger}c_{\bf R} +
\tilde{E}_{f {\bf R}} f_{\bf R}^{\dagger}f_{\bf R} +
\tilde{V}_{\bf R} \left(c_{\bf R}^{\dagger}f_{\bf R} + c.c.\right)\right)
\nonumber\\
 &-&   \sum_{\bf R}\sum_{{\bf \Delta} n.n.}
\left(t_c c_{\bf R + \Delta}^{\dagger}c_{\bf R} +
t_f f_{\bf R + \Delta}^{\dagger}f_{\bf R}\right)
\end{eqnarray}
where the effective one-particle parameters
\begin{eqnarray}
\tilde{E}_{c {\bf R}} &=& E_c + U \langle f_{\bf R}^{\dagger}f_{\bf R}\rangle
\nonumber \\
\tilde{E}_{f {\bf R}} &=& E_f + U \langle c_{\bf R}^{\dagger}c_{\bf R}\rangle
\nonumber \\
\tilde{V}_{\bf R} &=& - U P_{cf}^{\bf R} = - U \langle c_{\bf
R}^{\dagger}f_{\bf R}\rangle
\end{eqnarray}
have to be determined selfconsistently together with the chemical potential
$\mu$ for a given total number of electrons per site 
$n= \frac{1}{N} \sum_{\bf R} (\langle f_{\bf R}^{\dagger}f_{\bf R} \rangle 
+ \langle c_{\bf R}^{\dagger}c_{\bf R}\rangle)$. In this paper we study the
half filled case, i.e. $n=1$ electron per site.
$\tilde{V}_{\bf R}$ corresponds to
an effective, spontaneous hybridization between $f$- and $c$-electron states, 
which exists only if there is a nonvanishing spontaneous polarization
\begin{equation}
\label{eq:Pcf} 
P_{cf}^{\bf R} = \langle c_{\bf R}^{\dagger}f_{\bf R}\rangle = 
\langle f_{\bf R}^{\dagger}c_{\bf R}\rangle \neq 0
\end{equation}
Without loss of generality we assume here that $P_{cf}^{\bf R}$ and 
$\tilde{V}_{\bf R}$ can be chosen to be real. 

Concerning the position dependence of the effective one-particle parameters
$\tilde{E}_{c {\bf R}}, \tilde{E}_{f {\bf R}}, \tilde{V}_{\bf R}$
either a homgeneous, translational invariant solution, i.e. no $\bf
R$-dependence, is possible or an inhomogeneous solution with a periodic
modulation of the expectation values 
\begin{eqnarray}
n_{f{\bf R}} &=& \langle f_{\bf R}^{\dagger}f_{\bf R}\rangle = n_{f0} +
\frac{1}{2} m_f \cos({\bf Q\cdot R}) \nonumber \\
n_{c {\bf R}} &=& \langle c_{\bf R}^{\dagger}c_{\bf R}\rangle = n_{c0} +
\frac{1}{2} m_c \cos({\bf Q\cdot R}) \nonumber \\
P_{cf}^{\bf R} &=& \langle c_{\bf R}^{\dagger}f_{\bf R}\rangle = P_{cf0}
\cos({\bf Q\cdot R})
\label{Eq:modulation}
\end{eqnarray}
will be assumed. Therefore, Hartree-Fock solutions
with an additional ordered structure are possible, and the treatment allows
for the investigation of effects as phase separation and resulting
(structural) phase
transitions within the (unrestricted) HFA. This is particularly important
for investigations of the (E)FKM, as it is well known 
(from exact results available for two
\cite{KennedyLieb86,BrandtSchmidt} or infinite\cite{BrandtMielsch}
dimensions) that for half filling ($n = 1$) the chessboard phase forms the
ground state (i.e. an A- and B-sublattice structure with different $f-$ and
$c-$ electron occupations on the A- and B- sublattice). Though it is also
known (in particular from numerical results for the original FKM in two
dimensions\cite{WatsonLemanski95,LemanskiFreericks02}) that even more
complex and interesting ordered phases may exist (e.g. striped
phases\cite{LemanskiFreericks02}), we will restrict our investigation to the
mentioned chessboard phase. This means that we assume a bipartite lattice
which can be decomposed into an A- and B-sublattice 
 and allow for different expectation values (occupation
numbers) on the A-and B-sublattice (as in the case of antiferromagnetism).
Therefore, we restrict the ${\bf Q}$-vectors in Eq. \ref{Eq:modulation}
to the nesting vectors ${\bf Q} = \pi (1,1,\ldots)$ or $\cos({\bf QR}) = \pm
1$ and have the possibility of two different values $n_{c,A/B}$, $n_{f,A/B}$,
$P_{cf}^{A/B}$ for the expectation values depending on whether ${\bf R} \in A$
or $\in B$ sublattice. Then the $m_{c,f}$ introduced in Eq.
\ref{Eq:modulation} is already the charge density wave (CDW) order
parameter, and a non-vanishing $P_{cf}$ is the order paramater describing 
spontaneous polarization, i.e. electronic (anti-) ferroelectricity or an
excitonic insulator.

\section{Results}
\label{sec:results}

The selfconsistent solution  of the HFA equations always yields a
homogeneous, translational invariant solution without a spontaneous
polarization. Then the $f$- and $c$-bands are simply shifted by the amount
$U n_c$ and $U n_f$, respectively. When the two bands are sufficiently far
apart from each other, the lower one will be completely filled (i.e.
$n_{c/f} = 1$) and the other one is empty (i.e. $n_{f/c} = 0$) as in the
case of a conventional semiconductor two-band model. However, when the bands
overlap, also other HFA solutions are obtained, namely either solutions with a CDW
order parameter or solutions with a spontaneous polarization. The HFA
solutions with an additional symmetry breaking and order parameter have the
lower energy (compared to the homogeneous solution) and, therefore, describe
the better and more reliable approximation to the true ground state.   

\begin{figure}[!ht]
\begin{center}
\parbox{7cm}{
{\bf a)} \\
\includegraphics[scale=0.38]{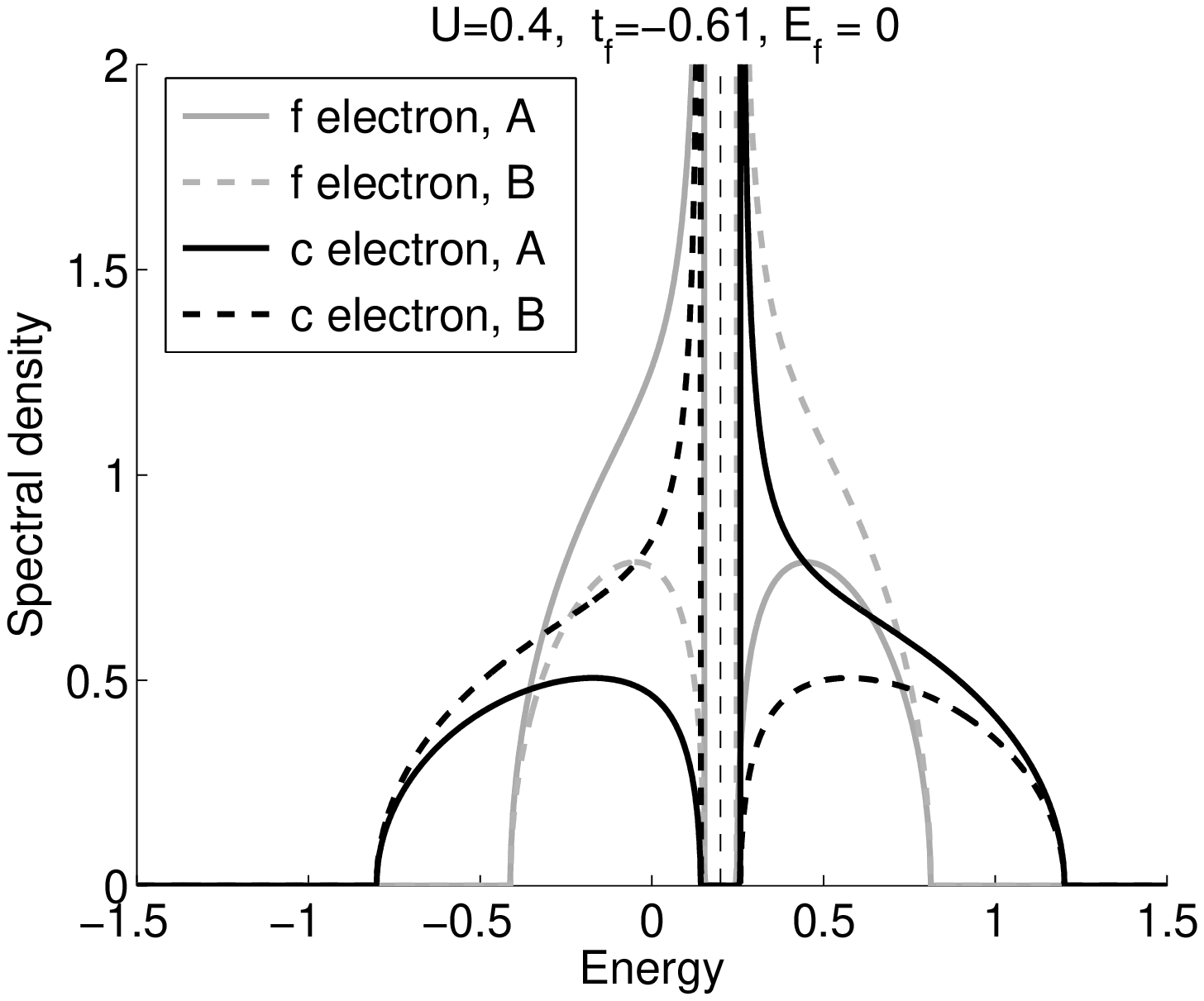}\\
{\bf b)} \\
\includegraphics[scale=0.38]{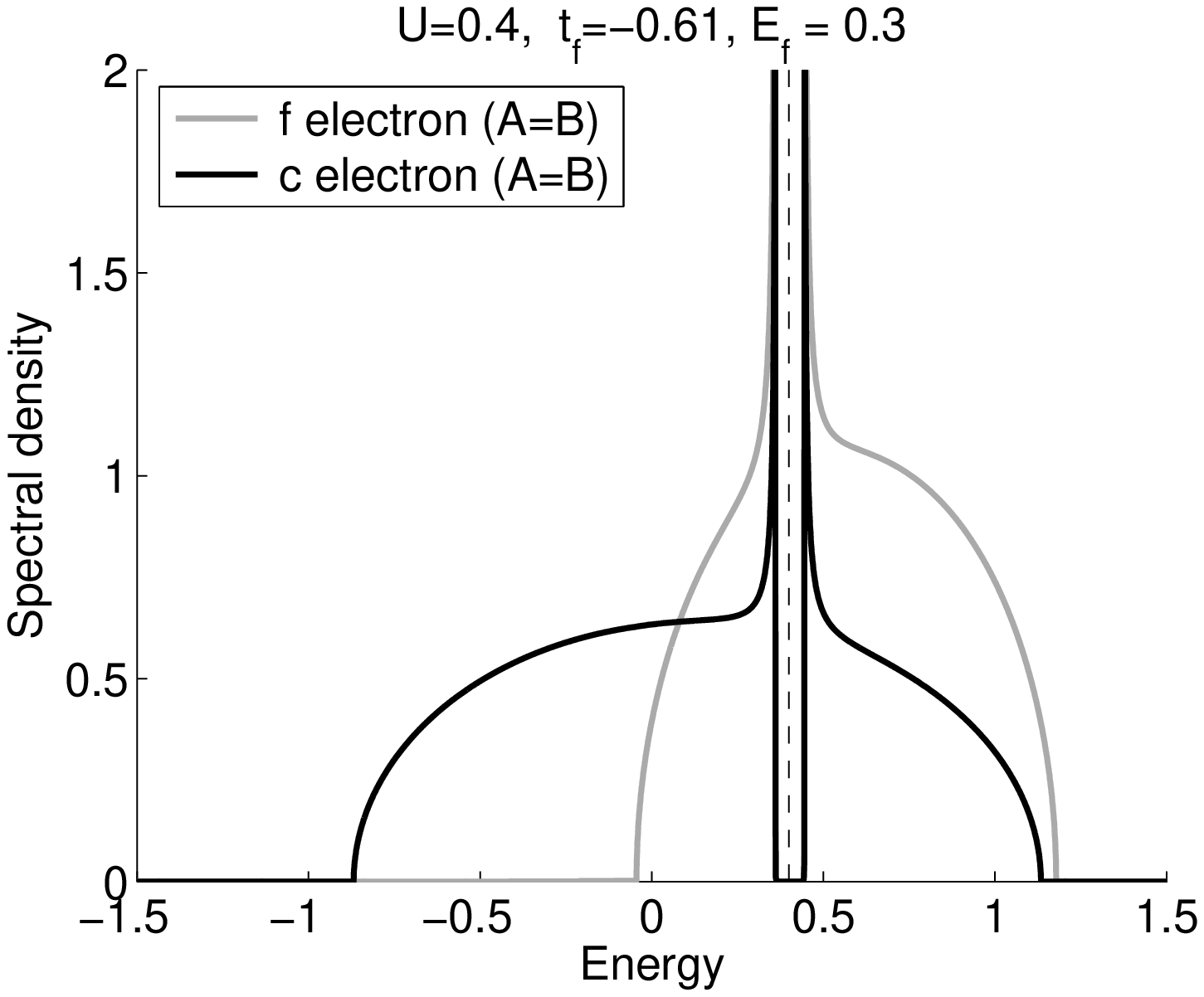}\\
{\bf c)} \\
\includegraphics[scale=0.38]{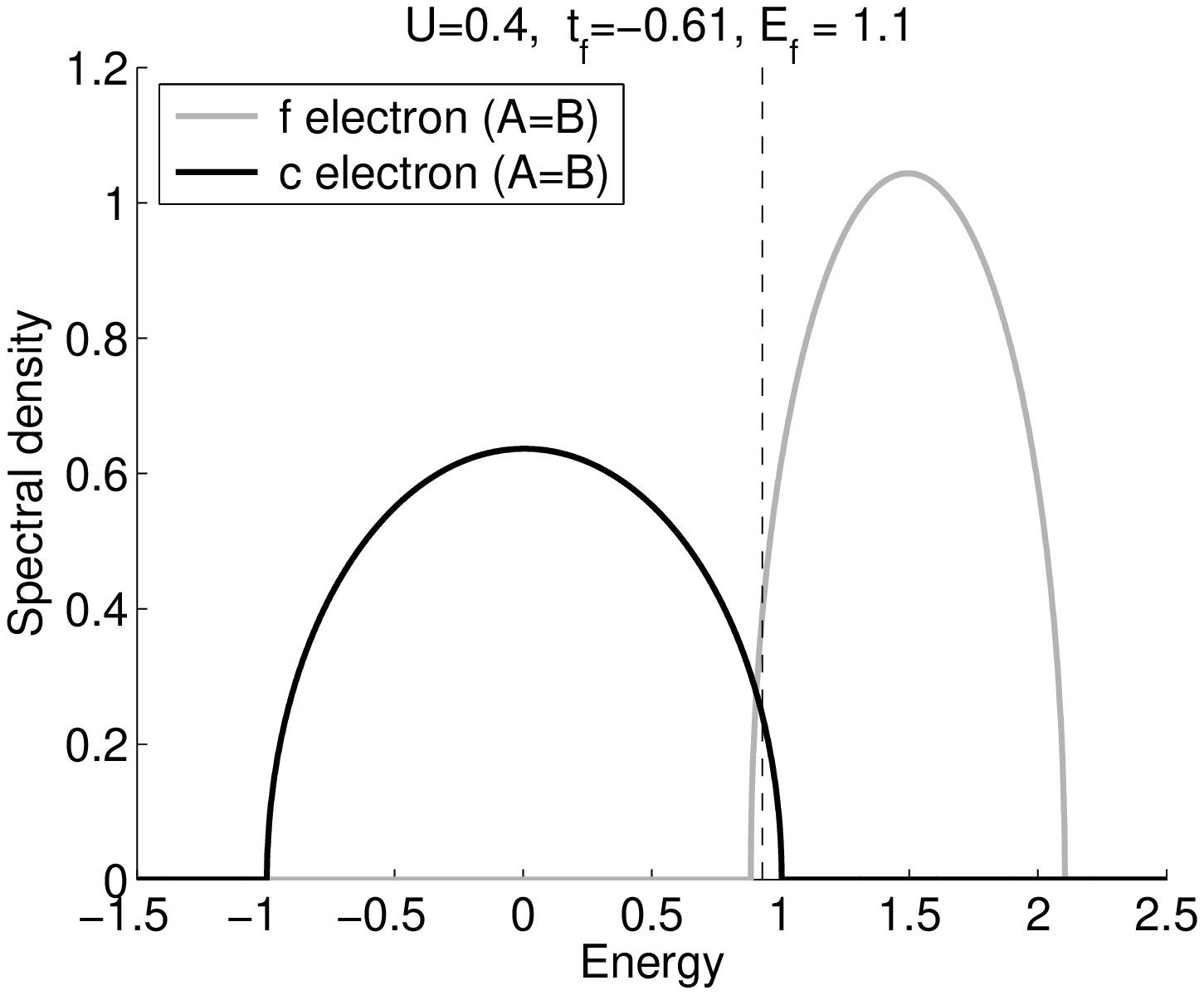}\\
{\bf d)} \\
\includegraphics[scale=0.38]{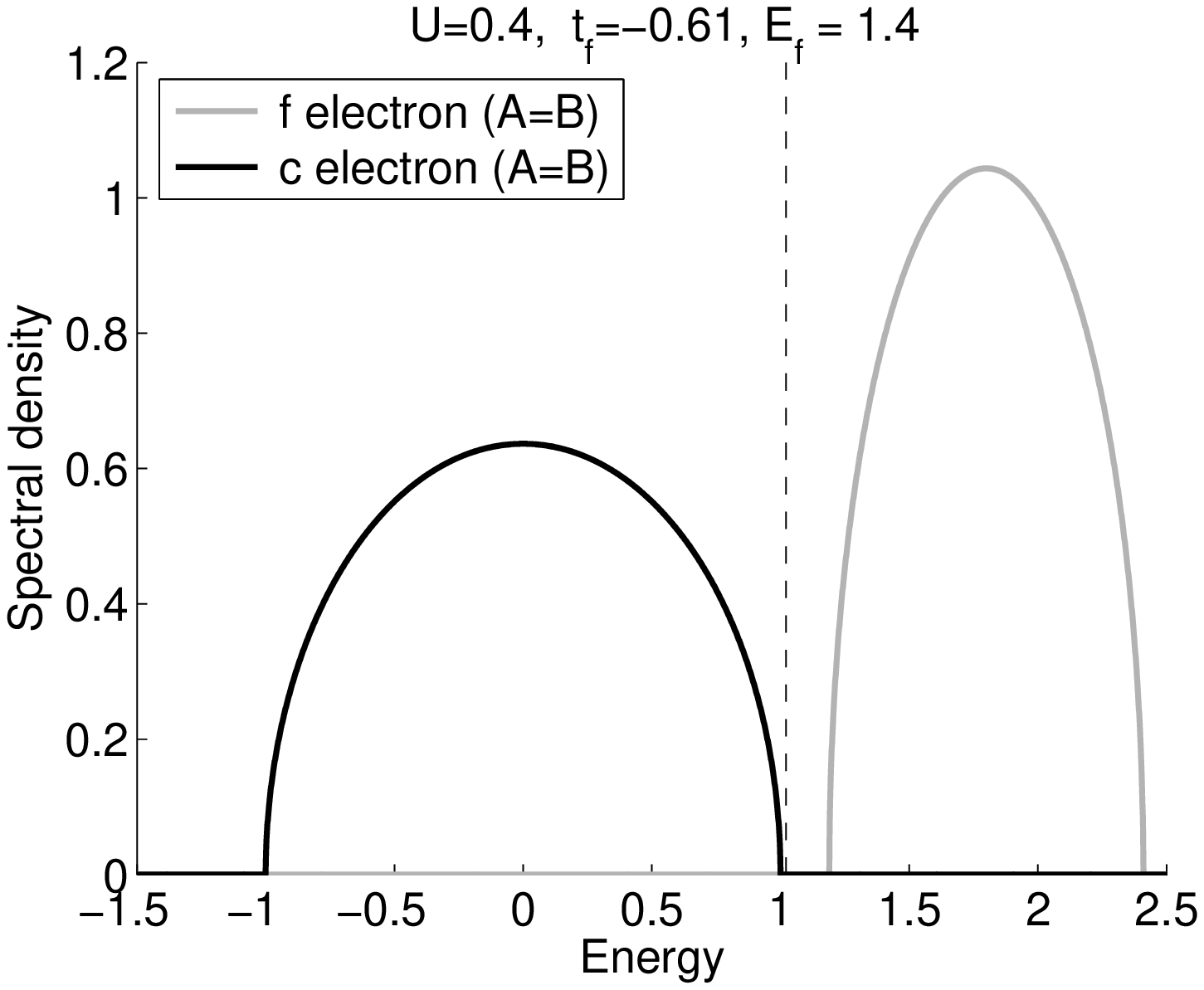}\\
\caption{$c$- and $f$-electron spectral function of the EFKM obtained within
the HFA for $U = 0.4, t_f = -0.61$, a temperature $T=0.002$ and different
$E_f$
}
\label{fig:spectralfunctions}
}
\end{center}
\end{figure}

To demonstrate the different types of HFA solutions obtained, several results
for the ($c$- and $f$-electron) spectral functions are shown in Fig.
\ref{fig:spectralfunctions} for $U=0.4$ and $t_f = - 0.61$ and four different
values of $E_f$. In the fully (particle-hole) symmetric case $E_f=0$, for
which a half filled $c$-band and a half filled $f$-band centered around
$U/2$ can be expected, a
charge density wave (CDW) solution is obtained as the most stable HFA solution. 
This
means that there are different occupation numbers $n_{cA} \neq n_{cB}$ and
$n_{fA} \neq n_{fB}$ on the different A- and B-sublattices. But in this
symmetric situation one has $n_{cA} = n_{fB}$ and vice versa, and $P_{cf}^{A/B}
= 0$, i.e. no spontaneous polarization. Because of the superstructure a
 CDW gap opens in the spectral functions, and as the Fermi energy $E_F
= U/2$ falls into this gap, an insulating solution with a CDW-gap is obtained. 
When the (center of the) $f$-band is shifted compared to the $c$-band
center, a different type of HF solution is obtained, namely one with a
non-vanishing $cf$-polarization or an effective hybridization, as
shown for $E_f=0.3$ in Fig.
\ref{fig:spectralfunctions} {\bf b)}. As the effective hybridization is site-diagonal
(local), a hybridization gap is formed and the chemical potential falls into
this hybridization gap. Therefore, for the total filling $n=1$ again an
insulating ground state is obtained, this time one with a hybridization gap.
This type of insulator is also termed ''excitonic
insulator''\cite{Kohn67,Keldysh64,Zittartz67}, and because of the
spontaneous $cf$-polarization $P_{cf}$ it is identical to the electronic ferroelectric
phase\cite{PortengenSham96,Batista02,Batistaetal04} and is sometimes also
called ''excitonic Bose-Einstein condensate'' (BEC) because of the
non-vanishing excitonic expectation value 
$\langle c_{\bf R}^\dagger f_{\bf R}\rangle$. 
There exists another type of solution as shown in 
Fig.\ref{fig:spectralfunctions} {\bf c)}, which is here obtained for $E_f=1.1$ and
corresponds to simply overlapping $c$- and $f$-bands. In this case both
bands are partially filled and the ground state is, therefore, metallic.
Because of the overlapping bands it is a semi-metal. In fact this
homogeneous, semi-metallic phase is obtained for all values of 
$E_f < 1 + |t_f| - U$ as a possible HFA-solution; for low temperature $T$ it
is usually not the energetically most favorable solution, but for
sufficiently high temperature $T > T_c$, where the possibly existing 
order parameters vanish,
it always becomes the stable phase. Finally, if $E_f$ is further shifted
upwards,  a situation is reached, 
where the two bands no longer overlap. Then
the lower band is totally filled and the upper band is empty, and $c-$ and
$f-$band are separated by a gap, because of which one has a conventional
band insulator. This situation is depicted in Fig.
\ref{fig:spectralfunctions} {\bf d)} for $E_f = 1.4$; it is obtained for $E_f+U-|t_f|
> 1$. 

\begin{figure}[!ht]
\begin{center}
\parbox{7cm}{
\includegraphics[scale=0.4]{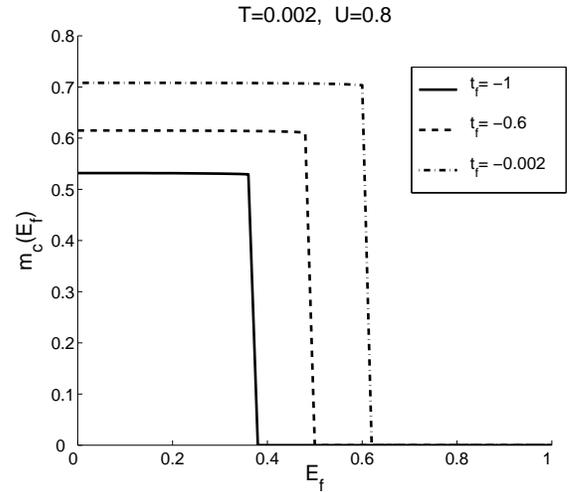}
\caption{\label{fig:CDW-orderparameter}CDW order parameter 
$m_c = |n_{cA} - n_{cB}|$ for $U=0.8, T=0.002$ as a function of $E_f$ for
different $t_f$}
}
\end{center}
\end{figure}

\begin{figure}[!ht]
\begin{center}
\parbox{7cm}{
\includegraphics[scale=0.4]{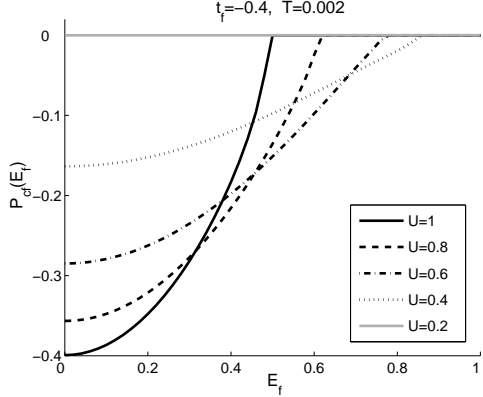}
\caption{\label{fig:FE-orderparameter} Ferroelectric order parameter
(spontaneous polarization) $P_{cf}$ for $t_f=-0.4, T=0.002$ as a function of
$E_f$ for different values of $U$}
}
\end{center}
\end{figure}

\begin{figure}[!ht]
\begin{center}
\parbox{7cm}{
\includegraphics[scale=0.4]{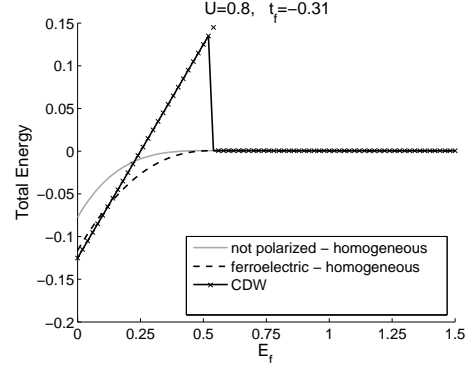}
\caption{\label{fig:totenergy}Total energy of the different HFA-solutions
for $U=0.8, t_f=-0.31$ as a function of $E_f$}
}
\end{center}
\end{figure}

The CDW phase and the ''electronic ferroelectric'' phase (with a spontaneous
polarization) are phases with a true symmetry breaking and an order
parameter. We have calculated these order parameters as a function of the
different parameters $U, t_f, E_f$ and as a function of temperature $T$. In
Fig. \ref{fig:CDW-orderparameter} the CDW order parameter $m_c$ is plotted
as a function of $E_f$ for fixed $U=0.8$ and temperature $T=0.002$ and three
values of $t_f$. Obviously, within the HFA a CDW is not only obtained for
the symmetric case $E_f=0$ but in a rather large interval of $E_f$-values
around 0. The CDW order parameter $m_c$ remains constant up to a critical
value of $E_f$ depending on $t_f$, where $m_c$ abruptly disappears. This
means, as a function of $E_f$ a first-order (quantum) phase transition is
obtained for the CDW order parameter $m_c$. 
The $E_f$-dependence of the ferroelectric order parameter $P_{cf}$ is 
shown in Fig.\ref{fig:FE-orderparameter} for fixed $t_f=-0.4$ and different
$U$. Obviously, a HFA solution with a non-vanishing $P_{cf} \neq 0$ is
obtained for all $E_f$ smaller than a critical value $E_{fc}(U)$, which
depends on the value of $U$. $P_{cf}(E_f)$ vanishes continuously when
approaching $E_{fc}$, i.e. one has a quantum phase transition of second
order from the ferroelectric to the homogeneous phase without symmetry
breaking. 
Altogether all three types of HFA solutions (CDW, spontaneous polarization
and homogeneous without symmetry breaking) exist for the same parameters, at
least for sufficiently small values of $|E_f|$. Then one has to calculate
the total energy to decide which HFA-solution is the most stable one and
comes closest to the true ground state. In Fig. \ref{fig:totenergy} we show
the dependence of the total HFA energy on $E_f$ for the three different
possible solutions for $U=0.8, t_f=-0.31$. One observes that for $E_f=0$ and
a small interval around 0 the CDW solution is energetically the most stable
one. The energy of the CDW state increases linearly with increasing $|E_f|$,
while the energies of the other HFA states  increase slower with increasing
$E_f$. At some value of $E_f$ the energy curves cross, and from
that critical value $E_{fc1}$ on the EFE solution with a spontaneous
polarization is energetically the most favorable one up to a second critical
$E_{fc2}$, where the FE solution merges into the homogeneous unpolarized HFA
solution. At this value $E_{fc2}$ the EFE order parameter vanishes
(continuously, cf. Fig. \ref{fig:FE-orderparameter}), i.e. there is a second
order quantum phase transition from the EFE to the homogeneous, unpolarized
solution. A CDW-solution exists also up to $E_{fc2}$, but from $E_{fc1}$ on
it is no longer the most stable HFA-soution. Therefore, there is a first
order quantum phase transition from the CDW solution to the EFE solution at
$E_{fc1}$.     

\begin{figure}[!ht]
\begin{center}
\parbox{7cm}{
\includegraphics[scale=0.4]{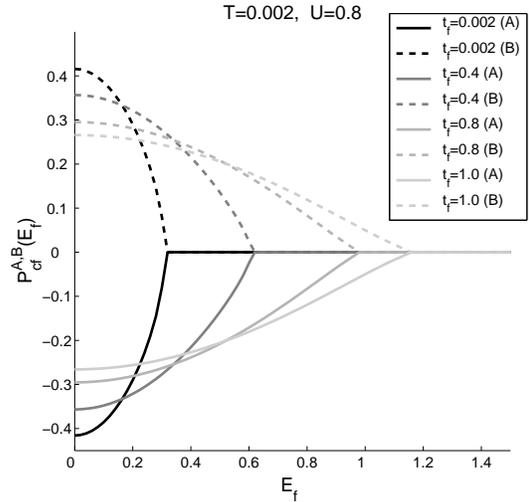}
\caption{\label{fig:AFE-orderparameter} Spontaneous polarization
$P_{cf}^{A/B}$ on the A- and B-sublattice as a function of $E_f$ for $U=0.8$
and different positive $t_f>0$}
}
\end{center}
\end{figure}

So far we have presented and discussed solutions for negative $t_f$. In the
case of positive $t_f > 0$ the CDW-solutions are completely unaffected; but
the EFE solution turns out to be unstable. Instead of this another phase with
a symmetry breaking and order parameter exists, namely a phase with an
AB-sublattice structure and different (opposite but equal in magnitude) 
polarizations on the A- and
B-sublattice, i.e. $P_{cf}^A = - P_{cf}^B \neq 0$. When the phase with a
homogeneous spontaneous polarization is termed ''electronic ferroelectric (EFE)'' phase,
the corresponding phase with non-vanishing but opposite polarizations on neighboring
sites must be called ''electronic anti-ferroelectric (EAFE)'' phase. For $U=0.8$ and
different $t_f > 0$ the polarizations 
on the two sublattices are shown in Fig. \ref{fig:AFE-orderparameter} as a
function of $E_f$. Obviously, except for the different sign the EAFE order
parameter behaves completely analogous as the EFE order parameter in the
case $t_f < 0$ (cf. Fig. \ref{fig:FE-orderparameter}). 

\begin{figure}[!ht]
\begin{center}
\parbox{7cm}{
\includegraphics[scale=0.35]{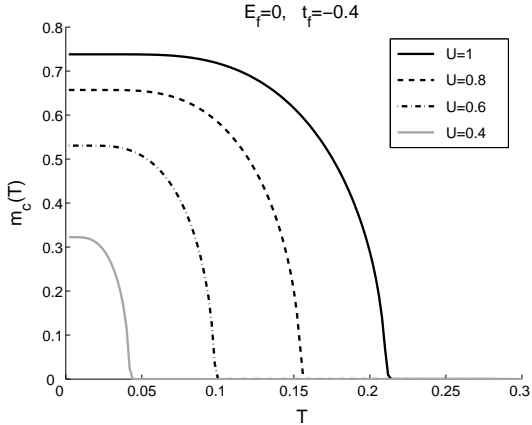}
\caption{\label{fig:T-dep-CDW-orderparameter} Temperature dependence of the
CDW order parameter $m_c$ for $E_f=0, t_f=-0.4$ and different $U$.}
}
\end{center}
\end{figure}

\begin{figure}[!ht]
\begin{center}
\parbox{7cm}{
\includegraphics[scale=0.35]{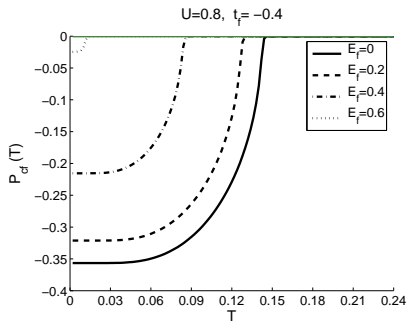}
\caption{\label{fig:T-dep-EFE-orderparameter} Temperature dependence of the
EFE order parameter $P_{cf}$ for $U=0.8, t_f=-0.4$ and different $E_f$.}
}
\end{center}
\end{figure}

The temperature ($T$-) dependence of the CDW order parameter $m_c$ is depicted in
Fig. \ref{fig:T-dep-CDW-orderparameter} for $E_f=0, t_f=-0.4$ and different
$U$. As it has to be expected from a HFA
treatment, the order parameter behaves mean-field like and vanishes
continuously (second order phase transition) 
at a critical temperature $T_c$ (with a critical index of
$\frac{1}{2}$); obviously $T_c$ increases with increasing $U$. Also
the $T$-dependence of the EFE order parameter $P_{cf}$,  shown in
Fig.\ref{fig:T-dep-EFE-orderparameter} for $U=0.8, t_f=-0.4$ and different
$E_f$ is mean-field like, and $P_{cf}$ vanishes at a $T_c$ depending on
$E_f$. 

\begin{figure}[!ht]
\begin{center}
\parbox{7cm}{
\includegraphics[scale=0.5]{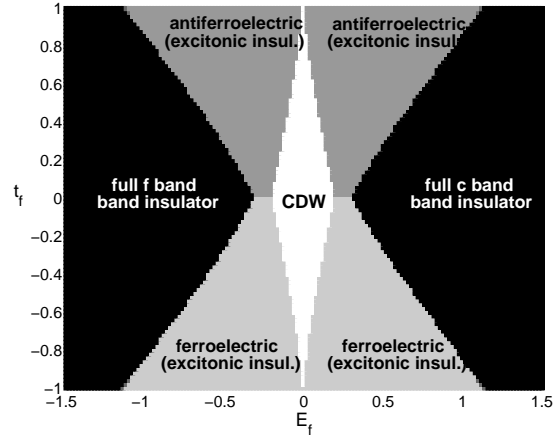}
\caption{\label{fig:phasediagU0.8} EFKM phase diagram in the $E_f-t_f$-plane
for $U=0.8$}
}
\end{center}
\end{figure}

The complete phase diagram obtained within the HFA is shown in Fig.
\ref{fig:phasediagU0.8} for fixed $U=0.8$ in the $E_f-t_f$-plane. We see
that -- besides the trivial phases of a completely filled $f$- or $c$-band
(for $n = 1$) -- the CDW phase and the EFE as well as the EAFE phase exist.
The CDW phase becomes broadest for $t_f = 0$, i.e. for the original
FKM\cite{FalicovKimball69}. Otherwise this line $t_f = 0$ is just the phase
boundary between EAFE and EFE phase. Therefore, just for the original FKM,
i.e. for $t_f=0$, no spontaneous polarization and no (anti-)ferroelectricity
has to be expected. This is in accordance with the fact that for the
original FKM the $f$-occupation operator $f_{\bf R}^\dagger f_{\bf R}$ at each site
is an exactly conserved quantity (commuting with the Hamiltonian), 
i.e. one has a conservation law for each
lattice site $\bf R$. But this special case $t_f = 0$ is an unstable fixed
point: an arbitrarily small finite (positive or negative) $t_f \neq 0$ leads
to an EAFE or EFE phase with a symmetry breaking and a finite order parameter. 
Altogether this phase
diagram is in complete  agreement with the one obtained previously by
different methods in the
case of strong\cite{Batista02} and intermediate\cite{Batistaetal04}
coupling for one- and two-dimensional systems. 
The same kind of phase diagram is also obtained in the
weak-coupling limit, in which the HFA applied here is valid.

\section{Conclusion}
\label{sec:conclusion}

We have investigated the extended Falicov-Kimball model (EFKM), i.e. a
spinless two-band model with a finite band width (hopping) of both, the $c$- and the
$f$-electrons, and a short-ranged (site diagonal, i.e. Hubbard like) Coulomb
interaction $U$ between $f$- and $c$-electrons. We have considered the
weak-coupling limit of small $U$, in which case the (unrestricted)
Hartree-Fock approximation (HFA) becomes applicable. For a total number of
$n=1$ electron per site different ground states are obtained depending on
the parameters, namely either the (trivial)
cases of completely  filled $c$- or $f$-band, states with a non-vanishing, spontaneous
$c-f$-polarization $P_{cf}^{\bf R}$ and a CDW-state with an A-B-sublattice
structure and different $c$- and $f$-electron fillings on the sites of the
A- and B-sublattice as in the case of antiferromagnetism. In both cases
there is a symmetry breaking, and the spontaneous polarization $P_{cf}$ and the
sublattice ''magnetization'' (difference of the A- and B-sublattice
occupation numbers) $m_{c,f} = |n_{c,f}^A - n_{c,f}^B|$ are the order
parameters. The state with a spontaneous polarization (and thus
hybridization) corresponds to the
excitonic insulator phase discussed already about 40 years
ago\cite{Kohn67,Keldysh64,Zittartz67}. A spontaneous polarization is
equivalent to a non-vanishing excitonic expectation value $\langle c_{\bf
R}^\dagger f_{\bf R}\rangle$, and a ground state with an excitonic
expectation value is sometimes also interpreted as an excitonic Bose-Einstein
condensate.\cite{Batista02,Batistaetal04} It may be connected with a spontaneous 
electric dipole moment (per site); therefore,  the state with a
translational invariant ($\bf R$-independent) $P_{cf}$ can be interpreted as
''electronic ferroelectricity'' (EFE)\cite{PortengenSham96}, 
which is obtained for $t_f < 0$. But for $t_f > 0$ a state with alternating
(opposite) $P_{cf}^{\bf R}$ on the A- and B-sublattice is obtained, which
can analogously be termed and interpreted as ''electronic antiferroelectricity''
(EAFE)\cite{Batista02,Batistaetal04}. We have determined the phase diagram
in the $E_f-t_f$-plane, which agrees with results obtained previously by
different methods\cite{Batista02,Batistaetal04}.    

Our conclusions are the following:
\begin{itemize}
\item Previous results on the EFKM, in particular the phase diagram,
obtained in the strong coupling limit by means of a mapping on an xxz-spin
model\cite{Batista02} or in the case of intermediate $U$ and in two
dimensions by means of a constrained path Monte Carlo
method\cite{Batistaetal04}, have been confirmed. This phase diagram can  also
be obtained 
in the weak-coupling limit of small $U$ within a simple
(unconstrained) Hartree-Fock treatment (independent of the dimension).

\item  The excitonic insulator state exists for this kind of two-band model
for a wide range of the parameters; it is equivalent to the EFE-state (for
$t_f < 0$) or the EAFE-state for $t_f > 0$. It is also equivalent to the
BCS-state (if one performs a particle-hole transformation for one kind of
the electrons and thus comes to a model with an attractive interband
interaction).

\item  Around the symmetric case, i.e. for (almost) coinciding band centers 
of the $c$- and $f$-band, another state is energetically more stable, namely
the CDW-state with different and alternating $c$- and $f$-occupation on the
sites of an A- and B-sublattice. In the older work on excitonic
insulators\cite{Keldysh64,Kohn67,Zittartz67} and on
EFE\cite{PortengenSham96} it has obviously not been checked, if a more
stable CDW-phase exists. 

\item There is a first order (quantum) phase transition from the E(A)FE
state to the CDW state but a second order transition from the E(A)FE phase
to the homogeneous (translationally invariant) phase (without a symmetry
breaking).

\item The original FKM\cite{FalicovKimball69} with $t_f=0$ corresponds to
the phase boundary between EFE- and EAFE phase. Therefore, neither the EFE-
nor the EAFE phase is stable and only the CDW-phase (and probably more
complex ordered structures or phase separation away from the symmetric case)
are the ordered phases for the original FKM, in agreement with previous
conclusions\cite{Czycholl99,Farkasovsky99}. However an arbitrarily small finite
$t_f \neq 0$ leads either to the EFE- (for $t_f < 0$) or to the EAFE-state
(for $t_f > 0$), at least sufficiently away from the symmetric case, when the CDW-state
is no longer favorable.
\end{itemize}

\vspace{0.3cm}

\noindent
{\bf Acknowledgments:}

\noindent
This work has been supported by a grant from the\\
''Bereichsforschungskommission Natur-/Ingenieur-\\wissenschaften (BFK NaWi)''
of the University of Bremen (project 01/102/9). We thank Frithjof Anders and
Claas Grenzebach for stimulating discussions. 

\vspace{0.3cm}

\noindent
{\bf Note:}

\noindent
After completion (but before submission) of this work, which is an extract
from Claudia Schneider's thesis\cite{ClaudiaSchneider04}, we became aware of
two very recent and related papers on the EFKM\cite{Brydon08,Farkasovsky08}.
Ref. \cite{Brydon08} applies the slave-boson theory to an EFKM extended
aditionally by a one-particle hybridization term. 
In Ref. \cite{Farkasovsky08} also the HFA has been applied (to the EFKM
without hybridization and two- and 
three-dimensional tight-binding bands), and the results are in complete
agreement with our results (obtained for the semicircular model density of
states, i.e. for a Bethe lattice in the limit of infinite coordination
number).

\end{document}